\documentclass[conference]{IEEEtran}
\IEEEoverridecommandlockouts
\usepackage{amsmath,amsfonts}
\usepackage{array}
\usepackage{textcomp}
\usepackage{stfloats}
\usepackage{url}
\usepackage{verbatim}
\usepackage{graphicx}
\usepackage{cite}
\usepackage[numbers,compress]{natbib}
\usepackage{amssymb}
\usepackage{caption}
\usepackage{subcaption}
\usepackage{bm}
\usepackage{multirow}
\usepackage{tabularray}
\usepackage{booktabs}
\usepackage{tabularx}
\usepackage{xurl}
\usepackage{physics}
\usepackage{float}
\usepackage[usenames,dvipsnames]{xcolor}
\usepackage{etoolbox} 
\usepackage{listofitems}
\usepackage{chngpage}
\usepackage{color}
\usepackage{soul}
\usepackage{caption}
\usepackage{eurosym}
\usepackage{tablefootnote}
\usepackage[short]{optidef}
\usepackage{orcidlink}
\usepackage[table]{xcolor}
\usepackage{tikz}
\usepackage{threeparttable}
\usepackage[final,commandnameprefix=ifneeded]{changes}
\usetikzlibrary{calc, arrows.meta, intersections, patterns, positioning, shapes.misc, fadings, through,decorations.pathreplacing, snakes, shapes.geometric, arrows}

\newcolumntype{P}[1]{>{\centering\arraybackslash}p{#1}}
\DeclareFontFamily{U}{mathx}{\hyphenchar\font45}
\DeclareFontShape{U}{mathx}{m}{n}{
	<5> <6> <7> <8> <9> <10>
	<10.95> <12> <14.4> <17.28> <20.74> <24.88>
	mathx10
}{}
\DeclareFontSubstitution{U}{mathx}{m}{n}

\newcommand{\midsepremove}{\aboverulesep = 0mm \belowrulesep = 0mm}
\midsepremove
\newcommand{\midsepdefault}{\aboverulesep = 0.605mm \belowrulesep = 0.984mm}
\midsepdefault

\begin{document}

\def\BibTeX{{\rm B\kern-.05em{\sc i\kern-.025em b}\kern-.08em
    T\kern-.1667em\lower.7ex\hbox{E}\kern-.125emX}}

\title{A review of imbalance price forecasting in Europe: algorithms, metrics and the way forward%
\thanks{Hussain Kazmi acknowledges support from the CELSA PREDDICT project. Margarida Mascarenhas acknowledges support from 1S39625N.}}

\author{\IEEEauthorblockN{Arnaud Verstraeten}
\IEEEauthorblockA{\textit{ELECTA (KU Leuven) and Gridual} \\
Leuven, Belgium \\
arnaud.verstraeten@kuleuven.be \\
arnaud@gridual.ai}
\and
\IEEEauthorblockN{Maria Margarida Mascarenhas}
\IEEEauthorblockA{\textit{ELECTA (KU Leuven)} \\
Leuven, Belgium \\
margarida.mascarenhas@kuleuven.be}
\and
\IEEEauthorblockN{Hussain Kazmi}
\IEEEauthorblockA{\textit{ELECTA (KU Leuven)} \\
Leuven, Belgium \\
hussain.kazmi@kuleuven.be}}

\maketitle

\begin{abstract}
Renewable electricity generation has grown significantly across many European power systems, leading to a greener energy mix, but also additional complexity in balancing electricity supply and demand. Unexpected differences between forecasts and actual output can lead to fluctuations in the system imbalance, which causes volatile imbalance prices. Accurate imbalance price forecasts are crucial for market players to choose a strategic balancing position. In early works, most forecasting methods combined fundamental and statistical approaches, but currently there is a clear trend towards data-driven machine learning models. This review compares forecasting algorithms in European markets with a focus on methodology. We emphasize the importance of high-quality input data, including intraday information and per-minute system data. Next, we identify the need for a common benchmark to compare novel forecasting methods developed for different markets and time periods. Finally, we argue that forecasts should be evaluated in terms of both downstream value and accuracy.

\end{abstract}

\section{Introduction}
To abate harmful greenhouse gas emissions resulting from electricity generation with fossil fuels, many power systems have introduced sustainable alternatives, including wind and solar energy \cite{iea}. The generation profiles of these renewable technologies can be difficult to predict, as they depend on weather conditions, such as wind speed and solar irradiation. As the share of variable renewable technologies in the energy mix expands, the task of balancing electricity load and generation becomes more difficult \cite{goodarzi}. To achieve a system-wide balance, each individual balance responsible party (BRP) must balance its electricity offtake and injection. For each settlement period, the transmission system operator (TSO) calculates the imbalance of every BRP. Depending on the sign of the collective system imbalance, BRPs can be penalized or remunerated based on their imbalance volume and the imbalance price. Clearly, the imbalance price is an important financial incentive to ensure that market players adequately balance their portfolio. However, due to the growing complexity of the balancing task, imbalance prices are becoming more extreme and more volatile. Accurate imbalance price forecasts are therefore crucial to ensure that imbalance prices provide the right incentives to market participants, even when prices are fluctuating strongly. For the TSO, imbalance price forecasts are instrumental to encourage BRPs to balance their position in the right direction, and ultimately mitigate the system imbalance \cite{pavirani}. For BRPs, imbalance price predictions are necessary to avoid severe penalties, or even generate profit by choosing a strategic balancing position \cite{smets}.

In parallel, the broader fields of time series and electricity price forecasting are evolving rapidly, because of the strong progress in forecasting techniques based on machine learning \cite{masini}. For example, ensembles based on decision trees or deep neural networks frequently achieve state-of-the-art forecast performance \cite{deng, lago_2021, mascarenhas, mascarenhas_asynchronous}. 

As a result, the field of imbalance price forecasting is growing in importance and popularity. In this review, we aim to give a clear overview of this relatively young research field, addressing the most interesting research trends, and highlighting the most significant research gaps. In Section \ref{sec:markets}, we discuss the importance of the balancing mechanism in European-style electricity markets. Hereafter, we examine the characteristics of imbalance prices. Section \ref{sec:literature} gives an overview of forecasting approaches used in literature. The next section discusses evaluation metrics and benchmarking methods. Finally, we list the most relevant challenges and research gaps.

\section{Balancing in European electricity markets}\label{sec:markets}
To minimize the system imbalance, each individual BRP has to balance its own portfolio as precisely as possible. Electricity markets are an essential tool for BRPs to obtain a balanced position. Because electricity prices and trading needs vary strongly as the delivery time comes closer, European electricity markets operate with a strict temporal hierarchy. Months and weeks before delivery, market players can buy and sell energy in forward markets. For trading one day in advance, the day-ahead (DA) market is used. The Single Day-Ahead Coupling\footnote{\url{https://www.entsoe.eu/network_codes/cacm/implementation/sdac/}} (SDAC) mechanism couples different European markets, and has a gate closure time (GCT) at D-1 12:00 CET. For shorter delivery lead times, market participants can trade in the intraday (ID) market, which consists of auctions and continuous trading. Similar to SDAC, intraday auctions are coupled on a European level through the Single Intraday Coupling\footnote{\url{https://www.nemo-committee.eu/sidc}} (SIDC) initiative. Three auctions are held (IDA1, IDA2, and IDA3), respectively closing at D-1 15:00, D-1 22:00, and D 10:00. 

Continuous intraday trading allows for trading even closer to the time of delivery. The exact ID gate closure time varies between regions. The first two columns of Table \ref{tab:markets} show the GCT for a selection of European markets. In every region, energy can be traded one hour before delivery. In Belgium, France, and the Netherlands, trading is even possible until the moment right before delivery. This allows BRPs to balance their portfolios until the final moment, which benefits the general system imbalance. In addition, market players can also financially profit from short gate closure times, as they can adjust their imbalances in accordance with accurate imbalance price forecasts that are based on the most recent system data.  

After the ID market has closed, the TSO determines the system imbalances and activates the necessary reserves. Then, the imbalance price can be determined based on the incurred balancing cost. In a single imbalance price system, all BRPs receive the same imbalance price. In a dual scheme, there can be an additional incentive for market players that aggravate the system imbalance. Table \ref{tab:dual_ip} summarizes the dual price system. $\alpha_1$ and $\alpha_2$ represent the additional price incentives. $p_\downarrow$ and $p_\uparrow$ respectively correspond to the imbalance price in the case of downward and upward regulations. When the system is long, the imbalance price is typically low (and sometimes even negative), as there is a general excess in energy. On the other hand, when the system is in need of electricity, imbalance prices are higher. When $\alpha_1 = \alpha_2 = 0$, the scheme in Table \ref{tab:dual_ip} boils down to a single imbalance price system.

With a single imbalance price, the cost of balancing the system (including the remuneration of BRPs that help the system) is exactly covered by the penalties imposed on BRPs that aggravate the system imbalance. Hence, a single imbalance price is the correct incentive to ensure that the system is perfectly balanced \cite{entsoe_workshop}. In some regions however, TSOs prefer to discourage positions on the wrong side of the system imbalance even more by employing a dual pricing system. This approach leads to an increased operational security, but it does not accurately reflect the balancing cost. In addition, especially smaller players are affected by this system, hampering the development of a level playing field. Therefore, article 52.2.c of the European electricity balancing guideline (EBGL) specifies that all TSOs should employ a single imbalance price \cite{ebgl}. The third column of Table \ref{tab:markets} shows the imbalance price systems used in different European countries. Five of the nine considered regions make use of a dual pricing system, indicating potential for change in the years to come.

Another important aspect of settling imbalances is the imbalance settlement period (ISP). Historically, many countries used settlement periods of half an hour or an hour. With the recent rise of variable renewables, generation levels can vary significantly within the span of an hour, which increases the need to settle imbalances more frequently. Article 53.1 of the EBGL states that all TSOs should utilize a settlement period of 15 minutes \cite{ebgl}. The final column of Table \ref{tab:markets} shows the ISP for the same nine regions as before. The United Kingdom is the only country that does not use an ISP of 15 minutes, as they are not a member of the ENTSO-E.

\begin{table}
	\centering
	\caption{Dual imbalance price system. If $\alpha_1 = \alpha_2 = 0$, this system is equivalent to a single imbalance price scheme. Inspired by \cite{bottieau_2020}.}
	\begin{tabular}{P{1.5cm} P{2cm} P{2cm}}
		\toprule
		\multirow{2}{1.5cm}{\centering \textbf{BRP imbalance}} & \multicolumn{2}{c}{\textbf{System imbalance}} \\
		& Long & Short \\
        \specialrule{.4pt}{2pt}{0pt}
		\multirow{2}{*}{Long} & \cellcolor{red!20} TSO pays BRP & \cellcolor{green!20} TSO pays BRP \\
        & \cellcolor{red!20} $p_\downarrow - \alpha_1$ & \cellcolor{green!20} $p_\uparrow$ \\
		  \multirow{2}{*}{Short} & \cellcolor{green!20} BRP pays TSO & \cellcolor{red!20} BRP pays TSO \\
        & \cellcolor{green!20} $p_\downarrow$ & \cellcolor{red!20} $p_\uparrow + \alpha_2$ \\
		\specialrule{.8pt}{0pt}{2pt}
	\end{tabular}
	\label{tab:dual_ip}
\end{table}

\section{Characteristics of imbalance prices}
Imbalance prices inherently depend on the sign of the system imbalance. Due to unexpected events such as the variable output of solar or wind energy, the sign of the system imbalance can switch quickly, leading to extremely volatile imbalance prices. 

\begin{figure}
	\centering
	\includegraphics[width=0.9\linewidth]{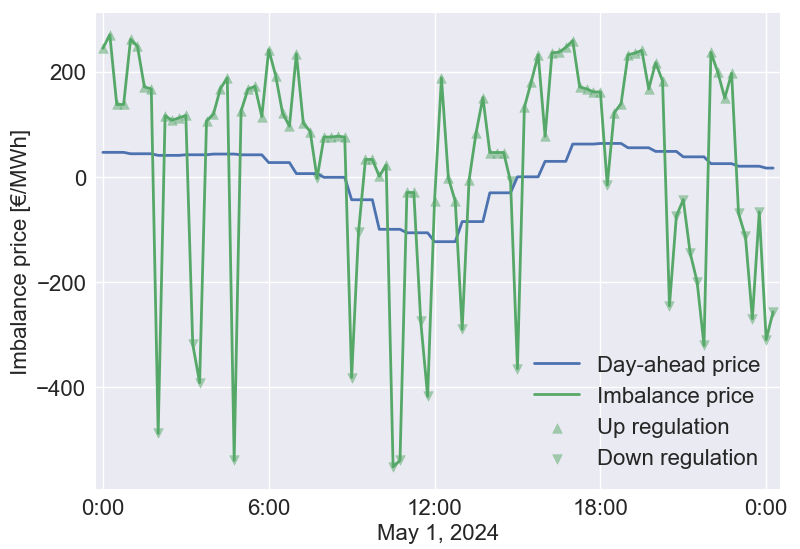}
	\caption{Belgian day-ahead and imbalance prices on a randomly selected day}
	\label{fig:ip_da}
\end{figure}

Fig. \ref{fig:ip_da} shows the Belgian day-ahead and imbalance prices on May 1, 2024. The day-ahead prices exhibit a typical pattern with low prices around noon and higher prices in the morning and evening peaks. Moreover, the price profile does not have any spikes or jumps. As a consequence, day-ahead prices can be accurately forecasted one day in advance. Imbalance prices on the other hand are much more volatile. On Fig. \ref{fig:ip_da}, the imbalance price follows the trend of the day-ahead price, but it constantly fluctuates due to changes in the direction of the regulation volume. Consequently, it is much harder to obtain an accurate imbalance price forecast.

\begin{figure}
	\centering
	\includegraphics[width=\linewidth]{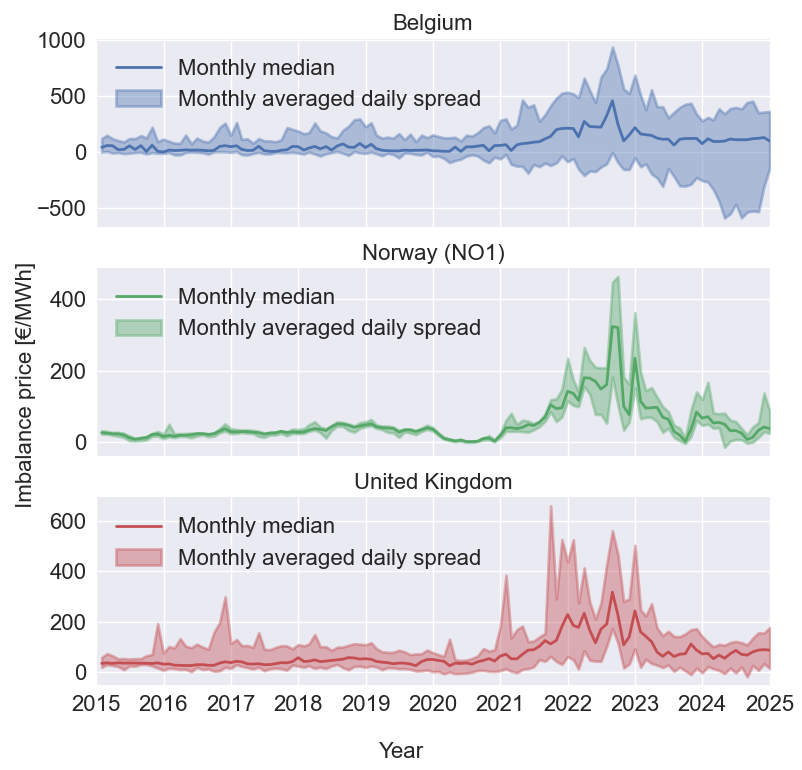}
	\caption{Monthly median and monthly averaged daily spread of the imbalance price for three different European regions. The daily spread is defined as the difference between the daily maximum and minimum of the imbalance price.}
	\label{fig:markets}
\end{figure}

In recent years, the volatility of imbalance prices has increased, mainly due to the growth of variable renewable generation technologies. Fig. \ref{fig:markets} shows the monthly spread in imbalance prices for three different European regions in the last ten years. In all three cases, we can distinguish two price trends. First, we observe that median imbalance prices have increased, indicating a higher cost for balancing the system. Secondly, in the last four years the imbalance price spread has strongly increased, as extreme imbalance prices occurred more frequently. Both of these trends underscore the relevance of accurate imbalance price forecasts.

To mitigate the balancing challenge that drives imbalance price volatility, the EBGL proposed the development of four different pan-European mechanisms for sharing balancing energy \cite{ebgl}. The TERRE, MARI, PICASSO, and IGCC platforms respectively cover the cross-border exchange of replacement reserves, manually activated reserves, automatically activated reserves, and imbalance netting, which is described in Articles 19--22 of the EBGL.

\section{Forecasting approaches}\label{sec:literature}
Imbalance price forecasting is a relatively small and young research field compared to day-ahead price forecasting. Because of the recent progress in forecasting techniques based on machine learning, the number of publications is growing rapidly. In this review, we focus on the methodology of different forecasting methods as opposed to other, more general reviews \cite{dinler, oconnor_review, singh}.

Table \ref{tab:literature} chronologically lists several research works that have imbalance price forecasting as their main objective. The first four columns provide the core attributes of each publication: the name(s) of the authors, the publication year, the considered market, and the main forecasting method(s). The seven remaining columns focus on methodology choices. These include the length of the horizon and lookback, the recalibration frequency, the size of the calibration window, the length of the test set, the type of forecast, and the evaluation metrics. The lookback corresponds with the lag of the time window that is used to include historical input features. The recalibration frequency is defined as the period between successive recalibrations of the forecasting model during the testing period. The calibration window is the length of the training set in each recalibration. The type of a forecast can be deterministic (D) or probabilistic (P).

On a high level, there are two main imbalance price forecasting approaches. The first approach breaks the prediction down in two steps. First, the system imbalance or balancing regime is predicted, after which this prediction is leveraged to make a forecast of the imbalance price. This approach leads to more fundamental models, as the imbalance price is determined based on the balancing regime. In the second approach on the other hand, the imbalance price is forecasted immediately without any additional steps, and the forecasting results are purely data-driven. This method is typically used in combination with powerful machine learning models such as transformers \cite{bottieau} or long short-term memory (LSTM) architectures \cite{backe, deng, smets2}. In early works, fundamental approaches were the most popular, but currently there is a clear trend towards complex machine learning architectures.

Another clear trend is that forecasting horizons have become shorter. Compared to early works in the Nordic market, horizons currently range from 15 minutes to several hours, rather than a full day or more. This trend is in line with an increased interest for trading close to real time. As was clear from the intraday gate closure times in Table \ref{tab:markets}, short imbalance price forecast horizons of a couple hours can already be informative for market players to choose strategic positions in intraday markets. In addition, short-term forecasts allow owners of flexible assets to perform implicit balancing in a risk-aware fashion \cite{smets}.

Next, it is also instructive to compare the lookback values used in different publications. With large lookback values of several hours or more, one can include more, potentially useful historical input information at the expense of a more complicated model. Table \ref{tab:literature} illustrates that there is no clear consensus on an optimal lookback value. Moreover, only a couple of publications assess the influence of the lookback parameter. Bottieau et al.\ \cite{bottieau} and Ganesh and Bunn \cite{ganesh_bunn} respectively find an optimal lookback value of 2 and 2.5 hours.

For the recalibration frequency, researchers seem to agree that it is not necessary to recalibrate the forecasting model for a test set shorter than two months. However, for a test set spanning a year or more, it is unclear how often a model should be recalibrated. In \cite{dol, narajewski}, the model is recalibrated daily, while in \cite{dumas} this is done on a monthly basis. In \cite{backe, bottieau}, the models are only trained once at the start of the testing period. 

Another important methodological choice is the size of the calibration window. The current literature seems to suggest that models should be trained with at least one year of data. Larger windows can lead to accuracy gains in exchange for a longer training time. So far, there exists no publication that attempts to optimize this tradeoff, or provides insight into the accuracy improvement with a larger calibration window.

The size of the test set is also an important choice that influences the assessment of the forecasting performance. In our opinion, the testing period should ideally span at least one year, in order to avoid any seasonal or short-term effects. However, in many works, a test set shorter than two months is used.

Further, the considered works can be grouped by the type of forecast that is produced. In general, most researchers opt for a probabilistic forecasting approach, which allows to quantify the uncertainty of the imbalance price prediction. 

\renewcommand{\arraystretch}{1.1}
\begin{table*}
	\scriptsize
	\centering
	\caption[Summary of imbalance price forecasting literature]{Summary of imbalance price forecasting literature. A backslash means that the considered characteristic does not apply to that case. A question mark indicates that the information is not specified.}
	\begin{adjustwidth}{-3.5cm}{-3.5cm}
	\centering
	\begin{tabular}{P{1.7cm} c c P{3cm} P{1.2cm} P{1.2cm} P{1.2cm} P{1.5cm} P{1.2cm} P{0.7cm} P{2.2cm}}
		\toprule
            \textbf{Paper} & \textbf{Year} & \textbf{Market} & \textbf{Method} & \textbf{Horizon} & \textbf{Look-back} & \textbf{Recal.\newline freq.} & \textbf{Calibration window} & \textbf{Test set} & \textbf{Type} & \textbf{Metrics} \\
		\midrule
		Olsson and Söder \cite{olsson_soder} & 2007 & Nordic & SARIMA +\newline Markov process & 24 h & 1 h & / & 1 y & / & / & Visual \\
    	Jaehnert \newline et al.\ \cite{jaehnert} & 2009 & Nordic & SARIMA +\newline long-term model & 48 h & 1 h & / & 5 y & 1 w & / & Visual \\
		Brolin and Söder \cite{brolin_soder} & 2010 & Nordic & ARMAX +\newline nonlinear model & 24 h & 2 h & / & 1 y & 2 m & / & Visual \\
		Kl\ae boe \newline et al.\ \cite{klaeboe} & 2013 & Nordic & ARMA +\newline Markov process & 1 h,\newline 36 h & 2 h & / & 2 y 5 m & $\sim$3 m & D, P & MAE, uncond. cov. \\
		Jónsson \newline et al.\ \cite{jonsson} & 2014 & Nordic & Holt--Winters model & 36 h & 1 d, 1 w & / & 14 m & 2 y & D & RMSE, $R^2$ \\
        Dimoulkas \newline et al.\ \cite{dimoulkas} & 2016 & Nordic & Hidden Markov model & 1 h,\newline 36 h & / & 1 h,\newline 1 d & $\sim$3 m & $\sim$9 m & D & MAE, SMAPE \\
		Dumas \newline et al.\ \cite{dumas} & 2019 & Belgium & Two-step approach & 6 h & / & 1 m & \textgreater1 y & 1 y & P & NMAE, NRMSE, PLF, CRPS \\
		Bunn \newline et al.\ \cite{bunn} & 2020 & GB & Markov regime-switching model & 30 min & 1 h & 30 min & $\sim$41 d & $\sim$41 d & D & RMSE, $R^2$ \\
		Lucas \newline et al.\ \cite{lucas} & 2020 & GB & Gradient boosting, random forest & 24 h & / & / & $\sim$11 m & $\sim$1 m & D & MAE, MSE, $R^2$ \\
		Lima \newline et al.\ \cite{lima} & 2022 & GB & Dynamic Bayesian model & 30 min & 1 h & / & $\sim$3 y & $\sim$41 d & P & PLF, RMSE \\
		Browell and Gilbert \cite{browell_gilbert} & 2022 & GB & Kernel density estimation and linear regression & 5 h & ? & ? & 3 y 9 m & 3 m & P & MAE \\
		Narajewski \cite{narajewski} & 2022 & Germany & Lasso with bootstrap, gamlss, prob.\ NN & 30 min & 30 min & 1 d & 730 d & 539 d & P & CRPS, RMSE, MAE, $\tau$\%-cov \\
        Bottieau \newline et al.\ \cite{bottieau} & 2022 & Belgium & Transformer & 4 h & 6 h & / & 3 y & 1 y & P & PLF, CRPS, Winkler score, PINAW, PICP \\
		Ganesh and Bunn \cite{ganesh_bunn} & 2023 & GB & Various NNs and statistical methods & 30 min & 2.5 h \newline (FCNN) & / & $\sim$3 y & $\sim$41 d & P & PLF, RMSE, MAE, $R^2$ \\
        Backe \newline et al.\ \cite{backe} & 2023 & Nordic & LSTM & 10 h & 10 h & / & 6 y & 1 y & D & RMSE \\
		Deng \newline et al.\ \cite{deng} & 2024 & GB & SA--BiLSTM & 2 h & 24 h & / & $\sim$3 y & $\sim$41 d & D & MAE, MAPE, RMSE \\
		Makrides \newline et al.\ \cite{makrides} & 2024 & Finland & Random forest & 1 hour & ? & ? & $\sim$1 y 7 m & $\sim$5 m & D & $R^2$, nRMSE \\
        O'Connor \newline et al.\ \cite{oconnor_conformal} & 2024 & Ireland & k-NN, lasso AR, random forest, gradient boosting & 8 h & 48 h & 8 h & 1 y & ? & P & PLF, interval width, coverage, Winkler \\
        Pavirani \newline et al.\ \cite{pavirani} & 2024 & Belgium & C--VSN + MCTS & 1--15 min & 4 h & / & 3 y & 10 d & P & MAE \\
        Smets \newline et al.\ \cite{smets2} & 2025 & Belgium & LSTM & 2.5 h & 1 h & / & 1 y 10 m & 2 m & D & ESS profit \\
        Plakas \newline et al.\ \cite{plakas} & 2025 & Greece & Gradient boosting +\newline Markov process & 1 h & 1 h & / & $\sim$11 m & $\sim$1 m & P & RMSE, MAE, NMAE, CRPS, PLF \\
        Dol \newline et al.\ \cite{dol} & 2025 & Netherlands & Lasso AR, gradient boosting & 1 h & 24 h & 1 d & 1 y & 1 y & D, P & MAE, RMSE, CRPS, Winkler, coverage \\
		\bottomrule
	\end{tabular}
	\end{adjustwidth}
	\label{tab:literature}
\end{table*}

Finally, we can also compare forecasting methods in terms of the considered input features. Across research works, the most popular input features include lagged imbalance prices, the system imbalance, and the day-ahead price. For fundamental models, this information can already lead to good predictions, as the forecasting approach is built on the principles behind the actual imbalance price calculation. In the case of data-driven machine learning methods on the other hand, adding additional features can enhance the quality of the model. Currently, there is a trend to consider an increasing number of input features, such as power plant generation schedules, solar irradiation, wind and load forecasts, the past errors on these forecasts, import and export schedules, and intraday market results. Although these variables can definitely be related to imbalance price values, it is not clear whether this information actually increases forecast quality, as machine learning models tend to suffer from limited explainability. 

Furthermore, despite the large variety of considered covariates, some inputs remain underexplored. A couple of works have leveraged intraday market results, but typically this data is behind a paywall, so that many researchers do not have access to this information. Next, with the introduction of pan-European platforms for exchanging balancing energy, imbalance price forecasts could be improved by including information on cross-border energy flows. Another data stream that is not included in current literature is per-minute system information. Some TSOs publish updates of the imbalance price and system imbalance every minute, which can significantly improve forecast quality in the very near term. In combination with fast-responding energy storage systems such as batteries, more accurate forecasts within a settlement period can unlock untapped trading value.

\section{Evaluation and benchmarking}
There are two important choices to make when evaluating imbalance price forecasts. The first choice is the evaluation metric, the second is the forecasting method used for benchmarking the prediction.

The last column of Table \ref{tab:literature} lists the evaluation metrics used in imbalance price forecasting literature. For deterministic forecasts, the metrics used in literature are the mean absolute error (MAE), the normalized MAE (NMAE), the mean square error (MSE), the root MSE (RMSE), the normalized RMSE (NRMSE), the mean absolute percentage error (MAPE), the symmetric MAPE (SMAPE), the $R^2$ value, or the downstream profit obtained in a trading case study. To evaluate probabilistic forecasts, researchers have used the continuous ranked probability score (CRPS), the pinball loss function (PLF), the Winkler score, coverage metrics such as the prediction interval coverage probability (PICP), or interval width metrics such as the prediction interval normalized average width (PINAW). 

To benchmark the performance of a certain forecasting method, it is customary to compare it with a naive method in terms of one of the aforementioned evaluation metrics. In the case of imbalance price forecasting, there are two common naive methods. The first method is a persistence model, which uses the most recent historical values of the imbalance price to make a prediction. Another naive method is to use day-ahead or intraday prices as an imbalance price forecast \cite{bottieau, browell_gilbert, narajewski}. 

\section{Challenges and research gaps}
After this review, we identify three research areas that can be improved in future research. The first area of improvement are the considered input features, where we identify four opportunities. The first topic is the availability of high-quality input data. Many researchers make use of renewables forecasts, but the forecasts provided by TSOs often have a low accuracy \cite{kazmi_tao}. Second, the data from intraday markets could be very informative to predict imbalance prices, but access to this data typically requires a paid subscription. A third opportunity related to the model inputs is to gain insight into the individual influence of each input feature. Currently, most research works choose an input feature set based on intuition, and they rarely analyze the effect of including or removing features (with the notable exception of \cite{bottieau_xgb}). A fourth solution for the improvement of forecasting models is to make use of data with a granularity of one minute. Many TSOs update important features such as the system imbalance every minute, which could be valuable information for a forecast that is updated at the same frequency. 

The second area of improvement is the comparison of different models. This challenge manifests itself both within a single research work, and between different publications. In current literature, we see a trend towards complex machine learning models. These models often come with many methodology choices that need to be optimized, but most works do not focus on optimizing all of these parameters. In this way, it is often unclear whether a model is actually performing optimally. Between research papers, it is even more difficult to compare the performance of various models. In many works, novel forecasting methods are introduced with promising results. However, usually it is not possible to compare the results with other works, as the models are tested in different markets and time periods. In that regard, it would be very interesting to organize an imbalance price forecasting competition, in order to find out which approach is the most effective.

The third area of improvement is the downstream value of the forecast. When evaluating a range of forecasting models, researchers are often only focused on the value of a predefined accuracy metric. However, in the case of imbalance price forecasting, accuracy is not necessarily the most important aspect. The ultimate goal of a useful imbalance price forecast is to support market players in making informed trading decisions. In recent works such as \cite{oconnor_conformal, smets, dol}, there is a trend in the right direction, where forecasting models are also analyzed based on separate battery trading case studies. This approach allows the use of different evaluation metrics, such as the operational profit of the storage system, and the energy contributed to mitigate the system imbalance.

\section{Conclusion}
The rise of variable renewable energy production has increased the need for balancing resources. To encourage BRPs to balance their position more precisely, ENTSO-E has advised TSOs to use a single imbalance price system and an imbalance settlement period of 15 minutes. In combination with short intraday gate closure times, this allows BRPs to leverage imbalance price forecasts to choose a strategic position. Accurate forecasts are crucial in this process, as European imbalance prices have become more volatile and more extreme in the last four years. In recent literature, powerful machine learning models are the most popular choice. The forecast accuracy of these methods is strongly influenced by the considered input features and methodology choices, but at this point it is not clear which features and hyperparameters lead to optimal forecasting results. Furthermore, we see great potential in the use of rarely used data sources, such as intraday market data and TSO data with a per-minute granularity. Finally, most research works currently focus on imbalance price forecast accuracy. Future research should also compare different models in terms of the financial value for market players and the system operator. 

\appendix
\begin{table}[H]
	\centering
	\caption{Intraday gate closure time, imbalance price system, and imbalance settlement period for a selection of European markets}
	\begin{threeparttable}
		\begin{tabular}{p{1.5cm} P{1.6cm} P{1.6cm} P{1.6cm}}
			\toprule
			  \textbf{Region} & \textbf{Intraday GCT} & \textbf{Imbalance price system\tnote{e,f}} & \textbf{Imbalance settlement period\tnote{e,f}} \\
			\midrule
		    Belgium & 0 min.\tnote{a} & Single & 15 min. \\
	        France & 0 min.\tnote{a} & Dual & 15 min. \\
	        Germany & 5 min.\tnote{a} & Single & 15 min. \\
	        Italy & 60 min.\tnote{b} & Dual & 15 min. \\ 
	        Netherlands & 0 min.\tnote{a} & Dual & 15 min. \\
	        Nordics & 60 min.\tnote{c} & Single & 15 min. \\
	        Portugal & 60 min.\tnote{d} & Dual & 15 min. \\
	        Spain & 60 min.\tnote{d} & Dual & 15 min. \\
	        United\newline Kingdom & 15 min.\tnote{a} & Single & 30 min. \\
			\bottomrule
		\end{tabular}
		\begin{tablenotes}
			\small
			\item[a] \footnotesize{\url{https://www.epexspot.com/en/media/6}}
            \item[b] \footnotesize{\url{https://www.rse-web.it/wp-content/uploads/2024/02/04_MI-inglese.pdf}}
            \item[c] \footnotesize{\url{https://www.nordpoolgroup.com/en/trading/intraday-trading/}}
            \item[d] \footnotesize{\url{https://www.omie.es/en/mercado-de-electricidad}}
            \item[e] \footnotesize{\url{https://transparency.entsoe.eu/}}
            \item[f] \footnotesize{\url{https://www.elexon.co.uk/bsc/settlement/imbalance-pricing/}}
		\end{tablenotes}
	\end{threeparttable}
	\label{tab:markets}
\end{table}

\bibliographystyle{IEEEtran}


\bibliography{references}

\vfill

\end{document}